\documentclass[twocolumn,prl,amsmath,amssymb,superscriptaddress, longbibliography]{revtex4-2}
\usepackage{bbm}
\usepackage{amssymb}
\usepackage{amsmath}
\usepackage{graphicx}
\usepackage{subfigure}
\usepackage{natbib}
\usepackage{epstopdf}
\usepackage{epsfig}
\usepackage{amsfonts}
\usepackage{mathrsfs}
\usepackage{CJK}

\usepackage{xcolor}
\usepackage[toc,page,title,titletoc,header]{appendix}

\begin{document}
\title{Photon Anomalous Blockade in Waveguide Cavity QED with Atomic Mirrors }
\author{Yang Xue}
\affiliation{School of integrated circuits, Tsinghua University, Beijing 100084, China}
\author{Yue Chang}
\affiliation{Beijing Automation Control Equipment Institute, Beijing 100074, China}
\affiliation{Quantum Technology R$\&$D Center of China Aerospace Science and Industry Corporation, Beijing 100074, China}
\author{Tao Shi}
\email{tshi@itp.ac.cn}
\affiliation{Institute of Theoretical Physics, Chinese Academy of Sciences, P.O. Box 2735, Beijing 100190, China}
\affiliation{CAS Center for Excellence in Topological Quantum Computation, University of Chinese Academy of Sciences, Beijing 100049, China}
\author{Yu-xi Liu}
\email{yuxiliu@mail.tsinghua.edu.cn}
\affiliation{School of integrated circuits, Tsinghua University, Beijing 100084, China}
\affiliation{Frontier Science Center for Quantum Information, Beijing, China}
\begin{abstract}
 Waveguide cavity quantum electrodynamics (QED) with atomic mirrors is a growing research area of quantum optics and can be applied to quantum information processing. We here study the photon statistics of output fields from a waveguide cavity QED system, in which the waveguide is coupled to quantized mirror atoms and one driven medium atom. Our results show that the photon blockade can occur even for a bad atom cavity with large dissipation and small coupling between the medium atom and the cavity, in contrast to the small dissipation and the strong coupling of the medium atom to the cavity field for the conventional photon blockade or the quantum interference for the unconventional photon blockade in the cavity QED system.  Utilizing both the master equation and scattering theories, we  derive the condition under which the photon blockade occurs in weakly driven systems. We find that such photon anomalous blockade is due to the quantum Zeno effect and is robust against variations of the medium atom's position within the cavity. Our study paves a way to exploit the photon blockade and single-photon devices via the waveguide cavity QED.
\end{abstract}
\maketitle

\textit{Introduction.---} Photon blockade is referred as a phenomenon that a single photon inside the cavity prevents from the injection of additional photons. The conventional photon blockade was originally found  in a driven cavity with weak dissipation and strong photon-photon interaction induced by the atom or other medium strongly coupled to the cavity~\cite{Tian1992PRA,Leonski1994PRA,Imamoglu1997PRL}.  The photon-photon interaction results in an anharmonic energy structure for such a cavity. Thus, the weak driving field with a given frequency cannot provide additional energy to resonantly excite higher energy level,  and the photons with the same frequency is blocked.  This conventional photon blockade can be used to convert classical light into non-classical one and serve as  single-photon turnstile devices~\cite{Imamoglu1997PRL} or excellent single-photon sources,  thus has potential applications in quantum optics and quantum information science~\cite{KnillN2001,HartmannNP2006,KokRMP2007,UmucallarPRL2012}.

The photon blockade has been experimentally demonstrated in various systems, e.g., cavity quantum electrodynamics (QED)  systems~\cite{Kim1999N,Smolyaninov2002PRL,Birnbaum2005N,Dayan2008S,Michler2000S,Faraon2008NP,Claudon2010NP,He2013NN,Madsen2014PRB,Gschrey2015NC,Somaschi2016NP,Dory2017PRA,Jia2018NP}, circuit QED systems~\cite{Hoffman2011PRL,LangPRL2011,Wang2016PRA,Liu2010PRA}, optomechanical resonators~\cite{Rabl,Xu2016PRA,Lemomde2016NC,Liao2013PRA} and others~\cite{Majumdar2012PRL,Majumdar2013PRB}.
The photon blockade can also occur when the nonlinear photon-photon interaction strength is weaker than the dissipation rate of the cavity~\cite{LiewPRL2010,BambaPRA2011}. This phenomenon, called as unconventional photon blockade (UPB)~\cite{FlayacPRA2017,SarmaPRA2017,ShenPRA2018,LemondePRA2014,Lu2025PRL}, arises from destructive quantum interference between transition paths, preventing simultaneous occupation of two-photon Fock states. The UPB has been experimentally observed in quantum dot cavity QED systems~\cite{SnijdersPRL2018} and coupled superconducting resonators~\cite{VanephPRL2018}.  Moreover, the photon blockade can also be achieved via quantum-reservoir engineering~\cite{Miranowicz2014PRA,Zhou2022PRA}. Universal photon blockade integrating single-photon resonance and quantum interference was recently also demonstrated via two-photon Jaynes-Cummings (JC) model~\cite{YCPPRL}.

\begin{figure}
\includegraphics[width=0.95\linewidth]{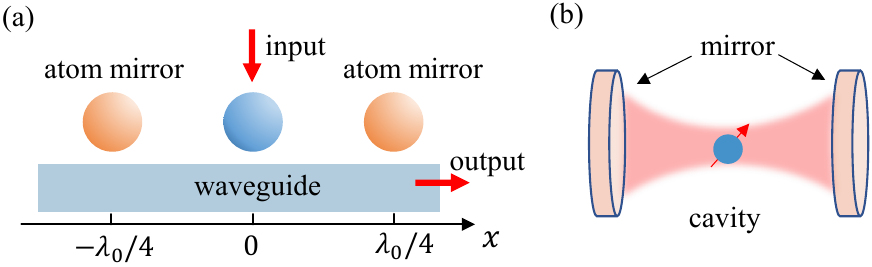}
\caption{(a) Schematics for the waveguide cavity QED system in which two quantized atom mirrors (brown) and a medium atom (blue) are coupled to the waveguide.  The medium atom located at $x_p=0$ is driven by a coherent light.  The mirror and  medium atoms are three identical two-level systems with the frequency $\omega_0$ corresponding to a wavelength $\lambda_0$ of photons. The waveguide cavity with the quantized atom mirrors is similar to the Fabry-Perot cavity with two parallel classical mirrors as schematically shown in (b).}\label{fig:fig1}	
\end{figure}

Recent advances in waveguide QED~\cite{Oleg} have motivated growing interest in waveguide cavity QED architecture. Different from the cavity QED system with classical  mirrors,  the waveguide cavity QED system has quantized mirrors formed by quantized atoms or artificial atoms~\cite{Zhou2008PRA,GongPRA2008,DongPRA2009,Guimond2016,CalajoPRL2019,Song2021OE,Regidor2021PRA,ZhouPRA2022,NiePRL2023}, and thus provides a new platform to study the light-matter interactions. The recent experiments have successfully demonstrated the waveguide cavity QED by coupling three transmon qubits to a microwave waveguide~\cite{Mirhosseini2019N}, where the middle transmon qubit acts as the medium atom and the cavity-like mode is formed by the dark state~\cite{Chang2012NJP} of other two transmon qubits acting as quantized mirrors. However, due to the excitations of atom mirrors, many phenomena in the waveguide cavity QED system are far from being thoroughly understood.

We here show a photon anomalous blockade in the waveguide cavity QED system~\cite{Mirhosseini2019N}. We find that the quality of the atom cavity abruptly becomes worse in two-excitation cases. Counterintuitively, such a bad cavity can result in photon blockade, which is neither from the strong coupling of the medium atom to the cavity~\cite{Imamoglu1997PRL} nor from the quantum interference~\cite{LiewPRL2010,BambaPRA2011}. Based on the master equation and scattering theories, we analytically derive the correlation function  in the waveguide cavity QED system~\cite{Shi2015PRA,Chang2016PRL}, and ultimately reveals the  mechanism of this  photon anomalous blockade, which is from the quantum Zeno effect.

\textit{Model of atom cavity with single-atom mirrors.---} As schematically shown in Fig.~\ref{fig:fig1}(a), we study a system that three identical two-level atoms with the frequency $\omega_0$ are coupled to a common waveguide. The middle one acts as a medium atom and other two are quantized mirrors forming into a cavity, which is nicknamed as the atom cavity. This is equivalent to a cavity QED system in Fig.~\ref{fig:fig1}(b) for a medium atom coupled to a microcavity. Hereafter, such coupled medium atom with mirror atoms via the waveguide is called as the waveguide cavity QED system.  We assume that  the medium atom is driven by a probe field with the frequency $\omega_d$ and the coupling strength $\epsilon$. In the rotating reference frame with $\omega_0$, the Hamiltonian of the system can be written as $H=H_w+H_d+H_{I}$ with
\begin{align}\label{eq:1}
H_w&=\sum_kk(\hat{r}_k^{\dagger}\hat{r}_k-\hat{l}_k^{\dagger}\hat{l}_k),\,\,\, H_d=\epsilon(\hat{\sigma}_pe^{i\Delta t}+\hat{\sigma}_p^{+}e^{-i\Delta t}),\nonumber\\
H_{I}&=\sum_{n=1,2,p}\sum_k\sqrt{\Gamma_n}\hat{\sigma}_n\left[\hat{r}_k^{\dagger} e^{-i(k+k_0)x_n}+\hat{l}_k^{\dagger} e^{i(k_0-k)x_n}\right]\nonumber\\
&\quad+\text{H.c.}.
\end{align}
 Here, $\Delta=\omega_d-\omega_0$ and $\hbar=1$. $H_w$ describes the free Hamiltonian of the waveguide photons, which propagate to the right or left around a central frequency $k_0=\omega_0$ corresponding to the photon annihilation operators $\hat{r}_k$ and $\hat{l}_k$, respectively. We set the photon propagating velocity $v_g=1$ for convenience. $H_d$ describes the interaction between the probe field and the medium atom. The ladder operator is defined as $\hat{\sigma}_n=|g_n\rangle\langle e_n|$ with the subscript $n=p$ or $n=1,\,2$ denoting the medium atom or two mirror atoms.  $|g_n\rangle$ and $|e_n\rangle$ denote the ground and excited states of the $n$th atom.  $H_{I}$ describes the coupling between the atoms and the waveguide, where $x_n$ represents the position of the $n$th atom. As shown in Fig.~\ref{fig:fig1}(a),  two mirror atoms locate at $x_1=-\lambda_0/4$ and $ x_2=\lambda_0/4$, respectively,  and the medium atom locates at $x_p=0$. The coupling strength between the $n$th atom and the waveguide is assumed as $\sqrt{\Gamma_n}$.

\begin{figure}
\includegraphics[width=0.9\linewidth]{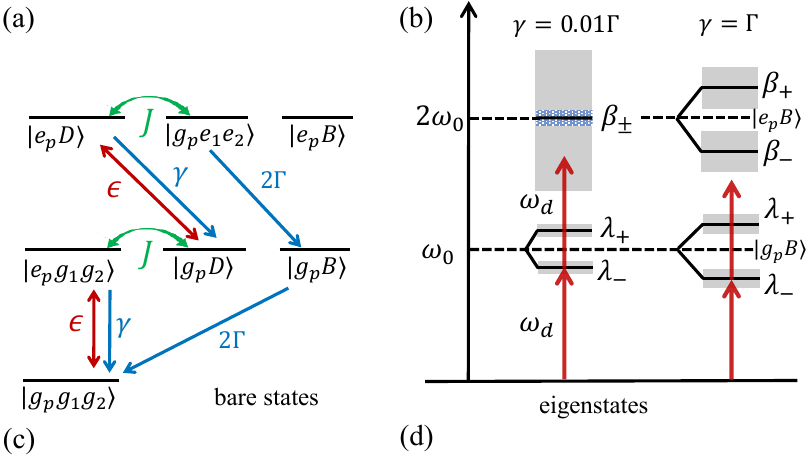}
\includegraphics[width=0.9\linewidth]{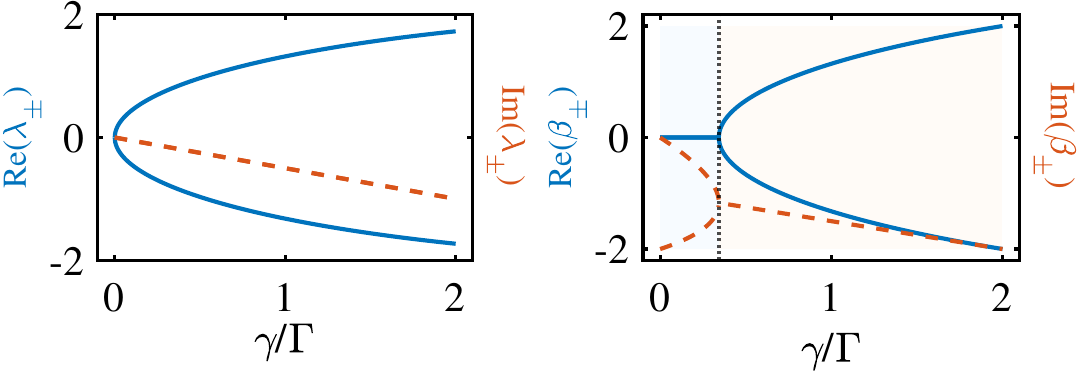}
\caption{(a) The transition diagram of the system when a probe field is applied to the medium atom. Each red line with two arrows denotes the probe field induced transitions between states. Each blue line with single-arrow means the decay.  Each green curve with two arrows means the waveguide induced mixing of two connected states. The coupling strength is $J=\sqrt{2\Gamma\gamma}$. Here, the states $|e_p, D\rangle=|e_p\rangle\otimes|D\rangle$ and $|g_p, D\rangle=|g_p\rangle\otimes|D\rangle$ include the branch of the dark state  $|D\rangle$, the states $|e_p, B\rangle=|e_p\rangle\otimes|B\rangle$ and $|g_p, B\rangle=|g_p\rangle\otimes|B\rangle$ include the branch of the bright state  $|B\rangle$. (b) Energy spectrum of the system for different decay rates $\gamma$ of the medium atom.  The lower energy splitting (solid line) represents the eigenvalues $\lambda_{\pm}$.  Similarly, the upper energy represents two eigenvalues $\beta_{\pm}$ with larger values of the imaginary parts. The gray and blue areas denote the imaginary parts of corresponding eigenstates. The dashed lines represent the energies $\omega_{0}$ and $2\omega_{0}$ corresponding to the states $|g_p,B\rangle$  and $|e_p,B\rangle$, respectively.  Red arrows denote the probe field frequency. (c) and (d) The real part (blue line) and imaginary part (orange line) of  eigenvalues $\lambda_{\pm}$ and $\beta_{\pm}$ in the single- and two- excitation subspace as a function of $\gamma$. The vertical line in (d) corresponds to $\gamma=2(3-2\sqrt{2})\Gamma$.}\label{fig:fig2}	
\end{figure}

Under Markovian approximation, the master equation of the reduced density matrix $\rho$ of three atoms without the probe field is
\begin{align}\label{eq:2}
\dot{\rho}=-i[H_c,\rho]+\mathcal{D}[\rho],
\end{align}
with $H_c=\sum_{mn}\sqrt{\Gamma_m\Gamma_n}\sin(k_0|x_m-x_n|)\sigma^+_m\sigma_n$. We here assume that the dissipations of all atoms are only from  the waveguide.  The terms with $n\neq m$ denote the coherent coupling between atoms induced by the waveguide,  however the terms with $n=m$ denote the Lamb shift caused by virtual photons in the waveguide. Besides, the waveguide also results in the dissipation of atoms described by a Lindblad form
$\mathcal{D}[\rho]=\sum_{mn}\gamma_{mn}(2\sigma_m^-\rho\sigma_n^+-\sigma_m^+\sigma_n^-\rho-\rho\sigma_m^+\sigma_n^-)$,
with $\gamma_{mn}=\sqrt{\Gamma_m\Gamma_n}\cos(k_0(x_m-x_n))$,  where $\gamma_{nm}$ with $n\neq m$ ($n=m$) denotes the waveguide-induced dissipative couplings between atoms (decay rates of the atoms). Thus,  an effective non-Hermitian Hamiltonian induced by the waveguide can be derived from Eq.~(\ref{eq:2}) as
\begin{align}\label{eq:3}
H_{\text{eff}}=-i\sum_{m,n}\sqrt{\Gamma_m\Gamma_n}e^{ik_0|x_n-x_m|}\sigma_m^{+}\sigma_n,
\end{align}
which can be rewritten as
\begin{align}\label{eq:4}
H_{\text{eff}}=-i\gamma\sigma_p^{+}\sigma_p-i2\Gamma S_B^{+}S_B+\sqrt{2\Gamma\gamma}(S_D^{+}\sigma_p+\sigma_p^{+}S_D)
\end{align}
for $x_p=0$, $x_1=-x_2=-\lambda_0/4$, where we assume $\Gamma_1=\Gamma_2=\Gamma$ and $\Gamma_p=\gamma$. Equation~(\ref{eq:4}) shows that the  collective operator $S^{+}_B=(\sigma_{1}^{+}-\sigma_{2}^{+})/\sqrt{2}$ is decoupled from that of the medium atom and is called as bright mode. It corresponds to a so-called bright state  $|B\rangle=S_B^+|g_1,g_2\rangle$, which is an eigenstate of the eigenvalue $-i2\Gamma$ for the Hamiltonian in Eq.~(\ref{eq:4}) of two mirror atoms in single-excitation space~\cite{Mirhosseini2019N}  when there is no medium atom. That is, this collective state decays to the waveguide at the rate $2\Gamma$. However,  the collective operator $S^{+}_D=(\sigma_{1}^{+}+\sigma_{2}^{+})/\sqrt{2}$ is coupled to the medium atom in the JC-like interaction and is called as dark mode. Thus, the dark mode mimics a cavity mode. Similarly,  a so-called dark state  $|D\rangle=S_D^+|g_1,g_2\rangle$ is an eigenstate with zero eigenvalue of the Hamiltonian in Eq.~(\ref{eq:4}) for two mirror atoms in single-excitation, i.e., this state does not directly decay to the waveguide. We note that the state $|e_1,e_2\rangle$ of two mirror atoms is also an eigenstate of the eigenvalue $-i2\Gamma$ for the Hamiltonian in Eq.~(\ref{eq:4}) when there is no medium atom, thus it is also a bright state. It is clear that the quantized mirror atoms possess different quantum states, which inevitably have  different effects on both the quality of the atom cavity and the cooperativity parameter~\cite{Mirhosseini2019N} of the waveguide cavity QED system.

\textit{Decays and driving-induced transitions.---} To show how the quantum states of the mirror atoms affect optical properties of the waveguide cavity QED system, we first study the decays of the eigenstates corresponding to the effective Hamiltonian in Eq.~(\ref{eq:4}) within single- or two-excitation subspace for two mirror and one medium atoms.  We know that the  bare states of these three atoms include the ground state $|g_p,g_1, g_2\rangle$, the single-excitation states $ |e_p, g_1, g_2\rangle, |g_p, D\rangle$, $ | g_p, B\rangle $, two-excitation states $|e_p,B\rangle, |e_p,D\rangle$, $ |g_p,e_1,e_2\rangle$, and three-excitation state $|e_p,e_1,e_2\rangle$ with the abbreviation, e.g., $|g_p,g_1, g_2\rangle=|g_p\rangle\otimes|g_1\rangle\otimes|g_2\rangle$,  $|e_p,D\rangle\equiv |e_p\rangle\otimes|D\rangle$. As shown in Fig.~\ref{fig:fig2}(a), some of these bare states are hybridized via the waveguide-induced coupling between atoms given in  the Hamiltonian in Eq.~(\ref{eq:4}).

In single-excitation subspace with the basis $|e_p, g_1, g_2\rangle$, $ |g_p, D\rangle$, and $| g_p, B\rangle$, using the non-Hermitian theory~\cite{Bender1998PRL,Ganainy2018NP,zdemir2019NM,Ashida2020AD}, we diagonalize the Hamiltonian in Eq.~(\ref{eq:4})  as $H_{\text{eff}} = \sum_n \lambda_n |\lambda_n^{R}\rangle \langle \lambda_n^{L}|$ via the biorthogonal basis satisfying the condition $\langle \lambda_n^{L} | \lambda_m^{R} \rangle = \delta_{nm}$ with  eigenvalues
\begin{align}
\lambda_{\pm}&=\frac{1}{2}\left(-i\gamma\pm\sqrt{\gamma(8\Gamma-\gamma)}\right),\;
\lambda_3=-i2\Gamma.
\end{align}
The eigenstates $|\lambda^{R}_{\pm}\rangle$  corresponding eigenvalues $\lambda_{\pm}$ are hybridized polariton states by the states  $|e_p, g_1, g_2\rangle$ and $ |g_p, D\rangle$,  with respective linewidths as shown in Fig.~\ref{fig:fig2}(b). It is clear that the real ${\rm Re}(\lambda_{\pm})$ and imaginary ${\rm Im}(\lambda_{\pm})$ parts of $\lambda_{\pm}$ denote the decay rates and frequency shifts induced by the waveguide. The state $| g_p, B\rangle$ corresponds to an eigenvalue $-i2\Gamma$ and is decoupled from other two single-excitation states.  In Fig.~\ref{fig:fig2}(c), we plot the eigenvalues $\lambda_{\pm}$ as functions of $\gamma$ and find that the decay rates of two corresponding eigenstates are identical, both are $\gamma/2$ within the range $\gamma \in [0,8\Gamma]$. Thus, similar to the JC model, the cooperativity and quality of the waveguide cavity QED system are determined by the coupling strength $\sqrt{2\Gamma\gamma}$ between the medium atom and the mirror atoms, decay rates of the medium atom and the states of the mirror atoms. However, the state $| g_p, B\rangle$  does not affect the quality of the single-excitation atom cavity.

If the system is in the  two-excitation subspace with  the basis $|e_p,B\rangle$, $|e_p,D\rangle$ and $|g_p,e_1,e_2\rangle$,  then the Hamiltonian in Eq.~(\ref{eq:4}) can be written as
\begin{align}
\begin{pmatrix}
-i2\Gamma-i\gamma & 0   &   0  \\
0 &             -i\gamma &   \sqrt{2\Gamma\gamma} \\
0 &         \sqrt{2\Gamma\gamma}   &   -i2\Gamma \\
\end{pmatrix},\label{eq:matrix}
\end{align}
which has the eigenvalues
\begin{align}
\beta_{\pm}&=\frac{-i}{2}(2\Gamma+\gamma\pm \sqrt{4\Gamma^2-12\gamma\Gamma+\gamma^2}),	\nonumber\\
\beta_3&=-i2\Gamma-i\gamma.
\end{align}
We find that the eigenstate corresponding to the eigenvalue $\beta_3=-i(2\Gamma+\gamma)$  is the basis state $|e_p,B\rangle$, which is  decoupled from other two basis states. The eigenvalues $\beta_{\pm}$ correspond to two eigenstates $|\beta^R_{\pm}\rangle$ as shown in Fig.~\ref{fig:fig2}(b), which hybridize two basis states $|e_p,D\rangle$ and $|g_p,e_1,e_2\rangle$. We find the imaginary parts of the eigenvalues $\beta_{\pm}$ are much larger than those of the eigenvalues $\lambda_{\pm}$. Especially, when the decay rate $\gamma$ of the medium atom satisfies $\gamma\in\left[0,2(3-2\sqrt{2})\Gamma\right]$, the eigenvalues $\beta_{\pm}$ are both pure imaginary number as shown in Fig.~\ref{fig:fig2}(d). Thus, the quality of the atomic cavity is worse for two-excitation than that for single-excitation,  because once the medium atom emits a single photon, the two-photon state $|e_p,D\rangle$ with the dark state $|D\rangle$ of the mirror atoms  is changed to the state $|g_p,e_1,e_2\rangle$, which  quickly decays  in the rate $2\Gamma$.

When a probe field is applied to the waveguide cavity QED system through the medium atom described by the Hamiltonian $H_{d}$ in Eq.~(\ref{eq:1}), transitions between the single- and two-excitation eigenstates with the selection rule occur as shown in Fig.~\ref{fig:fig2}(a).  We find that the probe field induced transitions between the state $|g_p,B\rangle$ and other states are forbidden, but the state $|g_p,B\rangle$ provides a fast dissipation channel for the two-excitation state $|g_p, e_1, e_2\rangle$. Furthermore, the two-excitation state $|e_p, B\rangle$ does not participate in the transitions to other states but provides the dissipation channel to the three-excitation state $|e_p,e_1,e_2\rangle$. As schematically shown in Fig.~\ref{fig:fig2}(b) for effective eigenvalue spectrum of the system with $\gamma=0.01\Gamma$, we find that different from the JC model with the strong coupling for the photon blockade~\cite{Imamoglu1997PRL}, the probe field with the frequency  $\omega_d=\omega_0 + \text{Re}(\lambda_{-})$ may induce transition from the singe-excitation  to the two-excitation subspace due to large linewidths of the two-excitation states corresponding to the eigenvalues $\beta_{\pm}$. In other words, when $2(\omega_0+\text{Re}(\lambda_{-}))$ is within the range $[2\omega_0+\text{Im}(\beta_+)/2,2\omega_0-\text{Im}(\beta_+)/2]$, the photon blockade cannot occur in the conventional cavity QED system~\cite{Imamoglu1997PRL}. However, as shown below, the photon is blockaded  in such a range in the waveguide cavity QED system.

\textit{Photon anomalous blockade.---}  To study photon blockade, we use second-order correlation function $g^{(2)}(\tau)=G^{(2)}(\tau)/|G^{(1)}(0)|^2$, to characterize the photon statistics of the output field with $G^{(1)}(0)=\langle\Psi| r_{\text{out}}^{\dagger}(t)r_{\text{out}}(t)|\Psi\rangle$ and
\begin{align}\label{eq:8}
G^{(2)}(\tau)=\langle \Psi|r^{\dagger}_{\text{out}}(t)r_{\text{out}}^\dagger(\tilde{\tau})r_{\text{out}}(\tilde{\tau})r_{\text{out}}(t)|\Psi\rangle,
\end{align}
with $\tilde{\tau}=t+\tau$  when the medium atom is driven. Here,  $r_{\text{out}}(t)=\sum_ke^{-ik (t-t_f)}r_k(t_f)/\sqrt{2\pi}$  and $|\Psi\rangle=|0\rangle\otimes|g_p,g_1, g_2\rangle$ with the vacuum state $|0\rangle$ of the waveguide~\cite{WallsQO,Sup}.

\begin{figure}
\includegraphics[width=0.98\linewidth]{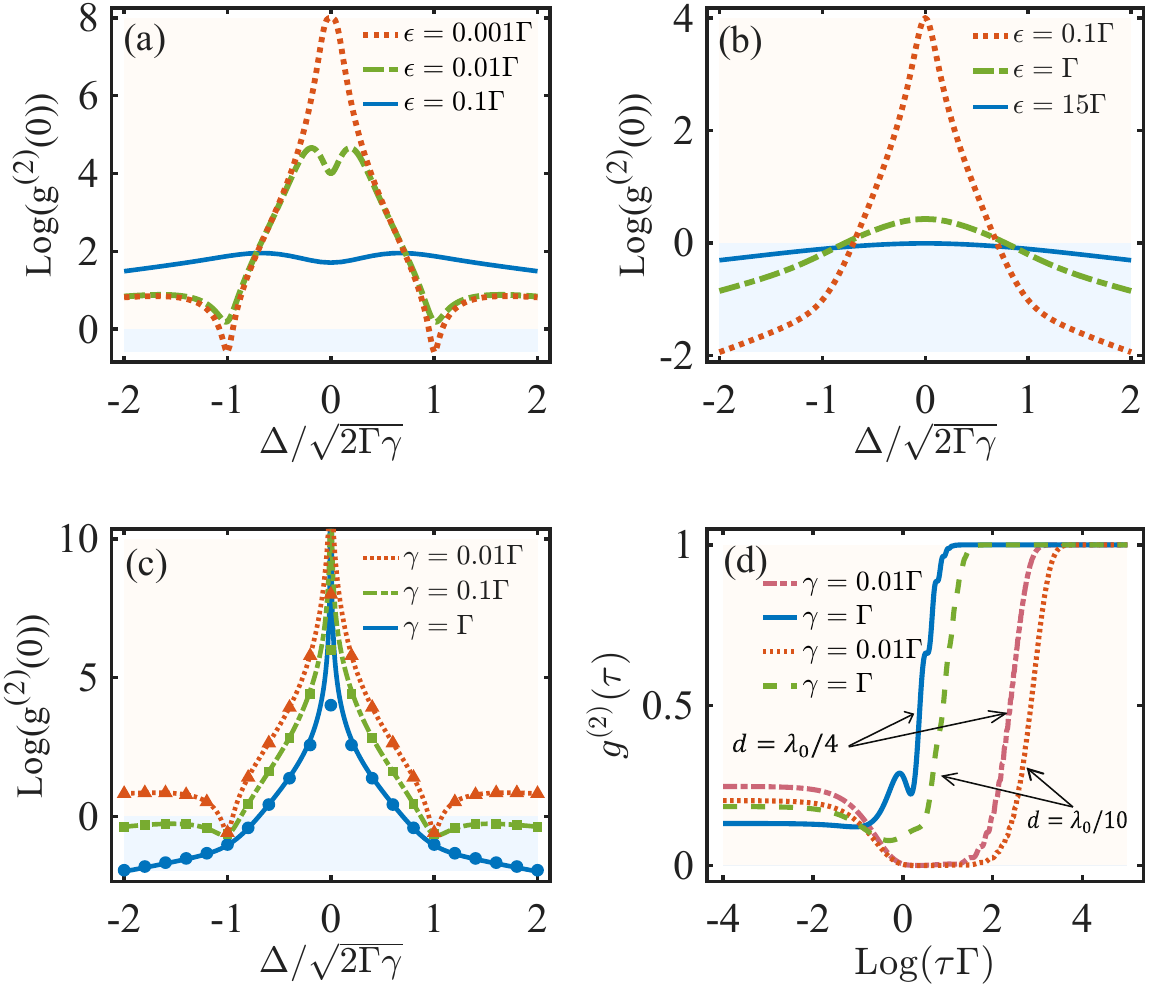}
\caption{(a) and (b) Correlation functions $g^{(2)}(0)$ as a function of the detuning $\Delta/\sqrt{2\Gamma\gamma}$ for different driving strengths $\epsilon$. The decay rate of the medium atom is $\gamma=0.01\Gamma$ in (a) and $\gamma=\Gamma$ in (b). (c) Correlation functions $g^{(2)}(0)$ obtained by the analytical results (blue solid, green dashed and red dotted curves) and direct numerical calculations by solving master equation (blue points, green squares, and red triangles) for different medium-atom decay rates $\gamma$. (d) Correlation functions $g^{(2)}(\tau)$ as a function of  time $\tau$ for distances $d=|x_p-x_1|=\lambda_0/4$ and $d=\lambda_0/10$, with separation between two mirror atoms fixed at $|x_2-x_1|=\lambda_0/2$ and detuning $\Delta=\text{Re}(\lambda_{-})\approx \sqrt{2\Gamma\gamma}$.}\label{fig:fig3}
\end{figure}

Under different driving strengths~\cite{Sup}, $g^{(2)}(0)$ is plotted as a function of the detuning $\Delta$ for different decay rates $\gamma$ of the medium atom. For $\gamma=0.01\Gamma$ as shown in Fig.~\ref{fig:fig3}(a), we find $g^{(2)}(0)<1$ at $\omega_d=\omega_{0}\pm {\rm Re(\lambda_{-})} \approx\omega_0\pm\sqrt{2\Gamma\gamma}$ when $\epsilon\leq\gamma$, indicating the photon antibunching and anomalous blockade. However, such photon blockade is very different from that in the JC model with strong coupling and weak dissipation~\cite{Birnbaum2005N}. Because the decay rate $2\Gamma+\gamma/2$ of the two-excitation state $|\beta^R_{+}\rangle$ is larger than the coupling strength $\sqrt{2\Gamma\gamma}$ between state $|e_p,D\rangle$ and $|g_p,e_1,e_2\rangle$, corresponding to the bad-cavity limit of the JC model, in which the photon blockade cannot occur. Moreover,  $g^{(2)}(0)$ exhibits only a single peak at $\Delta = 0$, in contrast to the three peaks observed in the JC model~\cite{Birnbaum2005N}. When the driving strength $\epsilon$ approaches and exceeds $\gamma$,  the antibunching effect gradually disappears. This is because a strong driving can excite the system from the single-excitation state to the two-excitation state within the lifetime $1/\gamma$ of the polaritons formed by the state of the single-excitation, thereby breaking single-photon blockade. However, for $\gamma = \Gamma$ as shown in Fig.~\ref{fig:fig3}(b), $g^{(2)}(0)$ decreases significantly with the increase of the detuning for a weak driving, indicating a strong antibunching. There exists an energy level splitting of $\sqrt{7}\Gamma$ between the two-photon eigenstates $|\beta^{R}_{\pm}\rangle$ when $\gamma=\Gamma$, identical to the splitting of $|\lambda^R_{\pm}\rangle$,  as shown in Fig.~\ref{fig:fig2}(b). This indicates that the observed blockade originates from a nonlinear level structure stronger than that of the JC model, in which such blockade is absent.

Let us  further analyze the mechanism of the photon anomalous blockade as shown in Fig.~\ref{fig:fig3}(a) for weak driving strength $\epsilon$, we analytically derive~\cite{Sup}
\begin{align}
g^{(2)}(0)=\frac{|\langle g|\tilde{\sigma}^2 \hat{G}_2\sigma_p^+\hat{G}_1\sigma_p^+|g\rangle|^2}{|\langle g|\tilde{\sigma}\hat{G}_1\sigma_p^+|g\rangle|^4}\label{eq:eq9},
\end{align}
with $|g\rangle=|g_p,g_1,g_2\rangle$ and $\tilde{\sigma}=\sum_n\sqrt{\Gamma_n}e^{ik_0x_n}\sigma_n$. For the numerator in Eq.~\eqref{eq:eq9},  from right to left, the first $\sigma_p^+$ represents that the system is excited via the medium atom. The off-shell Green's function $\hat{G}_1 = (\Delta - H^{(1)}_{\text{eff}})^{-1}$ describes the dynamics of the system  with the frequency $\Delta$ governed by the effective Hamiltonian $H^{(1)}_{\text{eff}}$ in the single-excitation subspace after the medium atom is excited. The second $\sigma_p^+$ depicts a transition from single-excitation to  two-excitation subspace via the medium atom, and the Green's function $\hat{G}_2 = (2\Delta - H^{(2)}_{\text{eff}})^{-1}$ describes the propagation of two excitations with the frequency $2\Delta$ governed by the effective Hamiltonian $H^{(2)}_{\text{eff}}$ in the two-excitation subspace.  Finally, the operator $\tilde{\sigma}^2$ accounts for the simultaneous annihilation of two excitations from the atomic ensemble.

In Fig.~\ref{fig:fig3}(c), $g^{(2)}(0)$ as a function of $\Delta$ is plotted for different $\gamma$ by using the analytical result in Eq.~(\ref{eq:eq9}), which agree well with those obtained via directly solving master equation under weak driving condition.
We now focus on the case where the driving frequency is resonant with the single-excitation mode, i.e., $\omega_d=\omega_{0}- \text{Re}(\lambda_{-})$, in which the output probability of single-photon is maximum~\cite{Sup}. For the resonant case, the initial excitation generated by the first $\sigma_p^+$ operator predominantly populates the long-lived single-excitation state $|\lambda^R_{-}\rangle$. The second $\sigma_p^+$ can in principle excite the system to the two-excitation state $|\beta^R_{+}\rangle$, however,  $|\beta^R_{+}\rangle$  is merely populated due to the large decay rate $\text{Im}(\beta_{+})$ compared to the strength $|\langle\beta_{+}^L|\sigma_p^+\hat{G}_1\sigma_p^+|g\rangle|$  of the two excitations. Thus, the larger decay rate of $|\beta^R_{+}\rangle$ relative to $|\lambda^R_{-}\rangle$ results in $g^{(2)}(0)<1$, a consequence of the quantum Zeno effect~\cite{Sun2023PRX,Misra1997,Itano1990PRA}, where the large decay rate $\text{Im}(\beta_+)$  implies an effective frequent measurements on the state $|\beta^R_+\rangle$, which is not occupied initially. This prevents the transition from $|\lambda^R_{-}\rangle$ to $|\beta^R_+\rangle$ and the photon anomalous blockade occurs.

In Fig.~\ref{fig:fig3}(d), the second-order correlation function $g^{(2)}(\tau)$ is further plotted as a function of the time delay $\tau$ rescaled by $\Gamma$ when the probe field resonantly excites the state $|\lambda_{-}^R\rangle$, i.e., $\Delta=\text{Re}(\lambda_{-})$. We find that the antibunching timescale between two photons is inversely proportional to the decay rate $\gamma$ of the medium atom. Moreover, the larger $\gamma$ leads to a smaller minimum value of $g^{(2)}(0)$. In addition, we find that  the antibunching timescale is gradually increased  when the distance $d = |x_p - x_1|$ between the medium atom and the left mirror atom is decreased from $\lambda_0/4$ to $\lambda_0/10$, while keeping the fixed decay rate $\gamma$ and the distance $|x_2 - x_1| = \lambda_0/2$ between two mirror atoms. This is because the spatial offset of the medium atom leads to the reduction of polariton decay rates in the single-excitation subspace.

\textit{N-atom mirrors.---} We now discuss the atom cavity, in which  each mirror consists of $N$ atoms ($N \geq 2$). Similar to the case of single-atom mirrors, we can define the collective operators of $2N$ mirror atoms via one bright mode and $2N-1$ dark modes~\cite{Sup} in the absence of the medium atom. In the single-excitation subspace, these bright and dark modes correspond to the bright and dark states, which are the eigenstates of the Hamiltonian of $2N$ mirror atoms with non-zero and  zero eigenvalues~\cite{Sup}. When the medium atom is coupled to the waveguide, we find that only one of dark modes is coupled to the medium atom, with an effective coupling strength of $\sqrt{2N\Gamma\gamma}$. Other $2N-2$ dark modes are decoupled from both the medium atom and the waveguide. Thus,  the coupled dark mode and the medium atom can possess two polariton states, which are similar to the case of single-atom mirrors.

In the two-excitation subspace,  in contrast to the case of single-atom mirror with only one bright state, $2N$ mirror atoms support $2N$ bright states and $C_{2N}^2-2N$ dark states. Similarly, the coupling of the  medium atom to the waveguide results in $C_{2N+1}^2$ polariton states, which hybridize the dark, bright and medium atom states.  When the medium atom is driven by the probe field, only three of these polariton states effectively participate in the dynamics. Two of polariton states, corresponding to the eigenvalues $2\omega_{0}\pm\delta$ have relatively small decay rates~\cite{Sup}. The transitions from the state of the single-excitation subspace to either one of these two polariton states cannot occur when the single-excitation state is resonantly driven,  and photon blockade occurs similar to the JC model. The third state with the energy $2\omega_0$ has a large linewidth,  the transitions from the states of the single-excitation subspace to this state are in principle allowed. However, the strong dissipation effectively suppresses such two-photon excitations and the photon blockade occurs via the quantum Zeno effect. Thus, we conclude that the photon blockade occur due to both JC-like nonlinear energy-level structure~\cite{Birnbaum2005N} and quantum Zeno effect~\cite{Misra1997,Itano1990PRA,Sun2023PRX} when the single mirror of the atom cavity consists of $N$ atoms ($N\geq2$).  We also find that the photons exhibit stronger antibunching when the number $N$ of the mirror atoms is increased.

\textit{Discussions and conclusions.---} The strong coupling  between three~\cite{Mirhosseini2019N} or eight~\cite{lin}  transmon qubits and the waveguide has recently been achieved. In particular, the decay rate of the medium qubit in the waveguide cavity QED system~\cite{Mirhosseini2019N} with single-atom mirrors can be engineered to $\gamma/2\pi = 1\,\text{MHz}$, while the decay rate of mirror atoms can be designed as $\Gamma/2\pi = 20\,\text{MHz}$ or $100\,\text{MHz}$. By weakly driving the medium qubit via a separate coplanar waveguide, the statistical property of output photons can be monitored via an extended Hanbury Brown-Twiss detection setup~\cite{Hamsen2017PRL}. We also mention that the atom cavity with $N$-atom mirrors (e.g., $N=3$) can also be realized via experiments~\cite{lin}. Thus, our study can be experimentally demonstrated via the waveguide QED systems~\cite{Xiang2013RMP,Gu2017PR} with the current state of the art.

In summary, we study the statistical properties of the output fields in a waveguide cavity QED system by weakly driving the medium atom. We find that the photon blockade can occur even in the bad-cavity limit, in which the coupling strength between the medium atom and the atom cavity is smaller than the decay rate. Compared to the JC model in the strong coupling limit~\cite{Birnbaum2005N}, the photon blockade is neither from anharmonic energy level structure~\cite{Imamoglu1997PRL} nor from the quantum interference~\cite{LiewPRL2010,BambaPRA2011}. This photon anomalous blockade results from the large decay for the eigenstates of the system in two-excitation subspace, which is equivalent to the quantum Zeno effect. We also find that the photon blockade is robust to the relative position of the medium atom to two mirror atoms with fixed distance $\lambda_0/2$. Moreover, when each mirror is formed by two or more atoms, the two-excitation eigenstates may hybridize the medium atom with the bright and dark states of the mirror atoms. In this case, the photon blockade occur due to both JC-like  anharmonic energy-level structure and quantum Zeno effect. We finally hope that our study can enrich and stimulate more explorations on the photon blockade and its applications in quantum information science.

{\it Acknowledgements.}
YXL was supported by the National Natural Science Foundation of China (Grants No. 92365209 and  No. 12374483) and Innovation Program for Quantum Science and Technology (Grant No. 2021ZD0300200).

\end{document}